\documentclass[a4paper]{jpconf}
\usepackage{graphicx}
\usepackage{braket}
\usepackage[T1]{fontenc}
\usepackage[ansinew]{inputenc}
\usepackage[english]{babel}
\usepackage{amsfonts}
\usepackage{amsmath}
\usepackage{array}
\usepackage{amsthm}
\usepackage{amssymb}
\begin{document}
\title{Quantum post-Newtonian theory for corpuscular Black Holes}

\author{Andrea Giugno}

\address{Arnold Sommerfeld Center, Ludwig-Maximilians-Universit\"at, Theresienstra{\ss}e 37,
80333 M\"unchen, Germany}

\ead{A.Giugno@physik.uni-muenchen.de}

\begin{abstract}
We discuss an effective theory for the quantum static gravitational potential
in spherical symmetry up to the first post-Newtonian correction.
We build a suitable Lagrangian from the weak field limit of the Einstein-Hilbert action
coupled to pressureless matter. Classical solutions of the field equation lead to the
correct post-Newtonian expansion. 
Furthermore, we portray the Newtonian results in a quantum framework by means of
a coherent quantum state, which is properly corrected to accomodate post-Newtonian 
corrections. These considerations provide a link between 
the corpuscular model of Dvali and Gomez and standard post-Newtonian gravity,
laying the foundations for future research.
\end{abstract}

\section{Introduction}
Newtonian theory describes gravity in terms of forces, allowing to well-define energy
and to provide the existence of a related scalar potential.
On the other hand, to specify these quantities in General Relativity (GR) proves to be a harder task, since it characterizes the gravitational field through the local
geometry of space-time~\cite{Weinberg}. Therefore, the notion of a scalar potential is not univocal for any observer.
In the post-Newtonian approximation, the local curvature is weak and the velocities of any
test particle are non-relativistic, which supports the derivation of an effective theory for the gravitational
potential generated by static and isotropic compact sources.
This formulation sees the approximated geodesic equation as a standard Newton's law, whose
potential is in turn determined by the Poisson equation.
We further enrich the picture by adding non-linearities in the quantum interpretation
of the gravitational potential~\cite{Casadio:2017cdv}, 
in the light of the results of Ref.~\cite{Casadio:2016zpl}.
It is important to stress out that this ``Newtonian-like" approach is one of the fundamental
ingredients, which allow to quantise gravitational degrees of freedom 
with standard methods~\cite{duff,donoghue,Casadio:2016fev,Casadio:2015qaq,Giusti:2017byx},
once some properties like spherical symmetry are implied.
The conceptual arena where this study is carried on, is the corpuscular model of gravity
brought to light by Dvali and Gomez~\cite{DvaliGomez,DvaliGomez2,DvaliGomez3,DvaliGomez4,DvaliSlava}. 
A black hole is naturally formed
by a large number of gravitons, which superimpose in the same quantum state and hence
realise a Bose-Einstein Condensate (BEC) stably on the verge 
of a quantum phase transition~\cite{flassig,becBH,becBH2,becBH3,becBH4,becBH5,becBH6}. 
In addition to that, the gravitons are expected to be marginally bound in their (gravitational)
confining potential well~\cite{qhbh,mueck,kerr}, whose size is set by the distinctive
Compton-de~Broglie wavelength $\lambda_{\rm G}\sim R_{\rm H}$, where
\begin{equation}
R_{\rm H}=2\,l_{\rm p}\,\frac{M}{m_{\rm p}}
\label{Rh}
\end{equation}
is the gravitational radius of the black hole of ADM mass $M$, and whose depth 
is directly proportional to the number $N_{\rm G}\sim M^2/m_{\rm p}^2$ 
of soft quanta~\cite{ruffini,Colpi:1986ye,Membrado:1989ke,Nieuwenhuizen:2008zz,Chavanis:2011cz}.
We mention here that we shall work in units of the Planck length $l_{\rm p}$ 
and mass $m_{\rm p}$, so that the Newton constant reads $G_{\rm N}=l_{\rm p}/ m_{\rm p}$, 
while $\hbar=l_{\rm p}\,m_{\rm p}$. The speed of light is instead normalised to $c=1$.
\par
Although the original outline~\cite{DvaliGomez4} only took the degrees of freedom
of gravity into consideration, especially when $R_{\rm H}$ is of astrophysical size~\cite{dvali,dvali2,kuhnel}, and substantially neglected the contributions of collapsing baryonic 
matter, the post-Newtonian approximation~\cite{Casadio:2016zpl} emerges in a 
simple fashion when they are properly taken into account.
The concept is straightforward: let us pretend to have $N$ baryons of rest mass $\mu$ 
very far apart, with their total ADM mass~\cite{adm} thus given by $M=N\,\mu$.
During the gravitational collapse, the baryons are enclosed in a sphere of radius $R$ and
possess a (negative) gravitational energy
$
U_{\rm BG}
\sim
N\,\mu\, V_{\rm N}
\sim
-\frac{l_{\rm p}\,M^2}{m_{\rm p}\,R}
$,
where $V_{\rm N}\sim -\frac{l_{\rm p}\,M}{m_{\rm p}\,R}$ is the associated Newtonian potential.
On a quantum mechanical perspective, the link with the classical potential $V_{\rm N}$ is achieved
thanks to the expectation value of a scalar field $\hat \Phi$,
$
V_{\rm N}
\sim
\bra{g}\hat \Phi\ket{g}
$, 
over a coherent state $\ket{g}$.
This feature entails that the graviton number $N_{\rm G}$ is determined by the normalisation
of the coherent state and follows Bekenstein's area law~\cite{bekenstein},
$
N_{\rm G}
\sim
R_{\rm H}^2/l_{\rm p}^2
$,
when $R=R_{\rm H}$.
Moreover, the Compton-de~Broglie length $\lambda_{\rm G}\sim R$ is recovered through
the assumption that it is the wave-length of almost the entire amount of gravitons.
Thus, the (negative) energy of any constituent yields consistently
$
\epsilon_{\rm G}
\sim
\frac{U_{\rm BG}}{N_{\rm G}}
\sim
-\frac{m_{\rm p}\,l_{\rm p}}{R} 
$
and the graviton self-interaction energy,
$
U_{\rm GG}(R)
\sim
N_{\rm G} \, \epsilon_{\rm G} \, \bra{g}\hat \Phi\ket{g}
\sim
\frac{l_{\rm p}^2\,M^3}{m_{\rm p}^2\,R^2}
$
reproduces the typical post-Newtonian correction.
It is easy to match this machinery with the standard knowledge. Considering indeed a star
of size $R\gg R_{\rm H}$ we get $U_{\rm GG}\ll \left|U_{\rm BG}\right|$, whereas
we recover the so-called ``maximal packing'' condition of Ref.~\cite{DvaliGomez},
\begin{equation}
U(R_{\rm H})
\equiv
U_{\rm BG}(R_{\rm H})+U_{\rm GG}(R_{\rm H})
\simeq
0
\ ,
\label{maxpack}
\end{equation}
when gravity is strongly coupled, \emph{i.e.} $R\simeq R_{\rm H}$.
\par
In the following, we first derive a consistent effective theory for a static and 
spherically symmetric potential by considering the Einstein-Hilbert action in the aforementioned
non-relativistic and weak field regimes. The inclusion of Next-to-Linear Order (NLO) terms 
provides classical results in compliance with the usual post-Newtonian expansion of 
the Schwarzschild metric. Furthermore, the quantum state of the soft scalar gravitons is
correctly represented by a coherent state, which establishes a link between a microscopic
description of gravity and the macroscopic geometry of space-time.
Therefore, such an outcome~\cite{Casadio:2017cdv} enriches and improves the conclusions of Ref.~\cite{Casadio:2016zpl}.
\section{Classical effective theory}
In order to build a correct effective theory, we first have to show how a real scalar field
can describe the post-Newtonian approximation of the weak field limit of 
Schwarzschild metric~\cite{Weinberg}.
The reader shall bear in mind, however, that this construction assumes that a specific 
reference frame has been chosen.
The starting point of our discussion is the Einstein-Hilbert action
\begin{equation}
S
=
\int \mathrm{d}^4 x\,\sqrt{-g}
\left(
-\frac{m_{\rm p}}{16\,\pi\,l_{\rm p}}\,\mathcal{R} +\mathcal{L}_{\rm M}
\right)
\ ,
\label{EHaction}
\end{equation}
coupled to the Lagrangian density $\mathcal{L}_{\rm M}$ that represents the ordinary
matter, which collapses and acts as a source for the graviational field. 
We labeled the Ricci scalar with $\mathcal{R}$.
As we already stressed out, in order to retrieve the well-known post-Newtonian result
the local curvature has to be small, which means that the metric can be expanded as
$g_{\mu\nu}=\eta_{\mu\nu}+h_{\mu\nu}$, where $|h_{\mu\nu}|\ll 1$, 
and $\eta_{\mu\nu}={\rm diag}(-1,+1,+1,+1)$.
Moreover, the characteristic velocity of the matter under consideration is 
many orders of magnitude smaller than the speed of light in the considered 
reference frame $x^\mu=(t,{\bf x})$.
We can safely describe the entire gravitational system through one relevant component of 
$g_{\mu\nu}$,  $h_{00}({\bf x})$, which is also time-independent.
The stress-energy tensor is hence completely specified by the non-relativistic energy density,
$
T_{\mu \nu}
\simeq
u_{\mu}\,u_{\nu}\,\rho({\bf x})
$,
where $u^\mu=\delta^\mu_0$ represents the four-velocity of the constituents of the source. 
Furthermore, this tensor allows to choose the matter Lagrangian simply as
$
\mathcal{L}_{\rm M}
\simeq
-\rho({\bf x})
$,
which is enough to our purpose, since we suppose that matter pressure 
is negligible~\cite{madsen88,brown1993} and we neglect the associated dynamics.
We can finally make the identification $h_{00}=-2\,V_{\rm N}$,
since the the Newtonian potential $V_{\rm N}$ is known to satisfy the Poisson equation
\begin{equation}
\triangle V_{\rm N}
=
4\,\pi\, \frac{l_{\rm p}}{m_{\rm p}}\, \rho
\ ,
\label{PoissonVN}
\end{equation}
while the de Donder gauge fixing takes the very simple form
$
\triangle h_{00}({\bf x})
=
-8\,\pi\, \frac{l_{\rm p}}{m_{\rm p}}\,\rho({\bf x})
$.
It is now straightforward to introduce an effective scalar field theory for the gravitational potential.
To do so, we first insert $V_{\rm N}$ in the Einsten-Hilbert action, and then we compute
the related Hamiltonian by making a Legendre transformation of the Lagrangian $L[V_{\rm N}]$. 
Then, we include non-linearities through the introduction of a self-gravitational source, 
defined as a the gravitational potential energy $U_{\rm N}$ per unit volume.
Leaving a more detailed analysis of such a construction to Ref.~\cite{Casadio:2017cdv} 
(and the Appendices thereof, in particular), here we only report the resulting 
Lagrangian
\begin{equation}
L[V]
=
-4\,\pi
\int_0^\infty 
r^2\,\mathrm{d} r
\left[
\frac{m_{\rm p}}{8\,\pi\,l_{\rm p}}
\left(1-4\,q_\Phi\,V\right)
\left(V'\right)^2
+q_{\rm B}\,V\,\rho
\left(1-2\,q_\Phi\,V\right)
\right]
\ .
\label{LagrV}
\end{equation}
The couplings $q_{\rm B}$ and $q_\Phi$ measure the strength of the interaction of $V$ 
with baryonic matter and the self-sourcing, respectively. They are rescaled in such a way to 
recover the known post-Newtonian expansion for $q_{\rm B}=q_\Phi=1$.
The Euler-Lagrange equation for $V$ is given by
\begin{equation}
\triangle V
=
4\,\pi\,q_{\rm B}\,\frac{l_{\rm p}}{m_{\rm p}}\,\rho
+2\,q_\Phi\left(1-4\,q_\Phi\,V\right)^{-1} \left(V'\right)^2
\ ,
\label{EOMVn}
\end{equation}
and it is obviously hard to solve analytically for a general source.
We will therefore expand the field $V$ up to first order in the coupling $q_\Phi$,
$
V(r)
=
V_{(0)}(r)
+
q_\Phi\, V_{(1)}(r)
\label{Vexp}
$,
and solve Eq.~\eqref{EOMVn} order by order.
In particular, for the leading order $V_{(0)}$ we have the analog 
of the Poisson Eq.~\eqref{PoissonVN}, while
\begin{equation}
\triangle V_{(1)}
=
2\left(V_{(0)}'\right)^2
\label{EOMVn1}
\end{equation} 
gives the correction at $\mathcal{O}(q_\Phi)$.
To linear order in $q_\Phi$, the on-shell Hamiltonian reads
\begin{equation}
H[V]
=
2\,\pi \int_0^\infty \mathrm{d} r \, r^2
\left[
q_{\rm B}\, \rho\, V \left(1-4\,q_\Phi \, V \right)
-q_\Phi\,\frac{3\,m_{\rm p}}{2\,\pi\,l_{\rm p}}\,V \left(V'^2 \right)
\right]
+
\mathcal{O}(q_\Phi^2)
\ .
\label{HamV}
\end{equation}
If we take \emph{e.g.} a matter distribution uniformly distributed 
within a sphere of radius $R$~\cite{Casadio:2017cdv}, 
not only the solution of the field Equation correctly reproduces
the potential that one expects in semiclassical gravity, 
but also the potential energy computed thanks to Eq.~\eqref{HamV} reads
\begin{equation}
U(R)
=
-q_{\rm B}^2\,\frac{3\,l_{\rm p}\,M^2}{5\,m_{\rm p}\,R}
+q_{\rm B}^3\,q_\Phi\,\frac{249\,l_{\rm p}^2\,M^3}{175\,m_{\rm p}^2\,R^2}
+\mathcal{O}(q_\Phi^2)
\ ,
\end{equation}
$M$ being again the ADM mass sourcing the graviational field up to order $q_\Phi$.
It is of paramount importance to notice that $U(R)$ vanishes when $R\simeq 1.2 R_{\rm H}$,
 realising therefore the ``maximal packing" condition~\eqref{maxpack}. 
\section{Quantum realisation}
In order to canonically quantise the theory, we consider the rescaled real scalar field
$
\Phi
=
\sqrt{\frac{m_{\rm p}}{l_{\rm p}}} \, V
$, coupled to the static source $
J_{\rm B}
=
4\,\pi\,\sqrt{\frac{l_{\rm p}}{m_{\rm p}}} \, \rho
$
and replace these new quantities in Eq.~\eqref{LagrV}.
We straightforwardly obtain the scalar field Lagrangian 
\begin{equation}
L[\Phi]
=
4\,\pi \int_0^\infty
r^2\, \mathrm{d} r 
\left[
\frac{1}{2}\,\Phi\,\Box \Phi
-q_{\rm B}\, J_{\rm B}\,\Phi
\left(1-2\,q_\Phi\,\sqrt{\frac{l_{\rm p}}{m_{\rm p}}} \,\Phi \right)
+2\, q_\Phi\, \sqrt{\frac{l_{\rm p}}{m_{\rm p}}}\,
\left(\partial_\mu \Phi \right)^2 \Phi
\right]
\ ,
\end{equation}
where we again assumed $\Phi=\Phi(t,r)$.
Let us look for a wave-function $\ket{g}$ which reproduces the classical solution
of the EOM of $\Phi$.
First, we examine the linear case, $q_\Phi=0$.
In terms of the new variables $\Phi$ and $J_{\rm B}$, it is analogous
to a Poisson Equation
$
\triangle\Phi_c(r)
=
q_{\rm B}\,J_{\rm B}(r)
\ ,
$
when the field is static as its own source.
Upon expanding the EOM on a base of spherically symmetric normal modes, one can solve
this Equation in momenta space, so that
\begin{equation}
\tilde\Phi_c(k)
=
-q_{\rm B}\,\frac{\tilde J_{\rm B}(k)}{k^2}
\ .
\end{equation}
A shift transformation of the ladder operators related to $\hat \Phi$ shows that the coherent state
$
\hat{a}_k
\ket{ g}
=
e^{i\,\gamma_k(t)}\,
g_k
\ket{g}
$
correctly reproduces the classical solution, 
$
\bra{g}
\hat \Phi(t,r)
\ket{g}
=
\Phi_c(r)
$,
with
\begin{equation}
g_k
=
-q_{\rm B}\,\frac{\tilde J_{\rm B}(k)}{\sqrt{2\,l_{\rm p}m_{\rm p}\,k^3}}
\ ,
\label{gk}
\end{equation}
and $\gamma_k(t)=-kt$.
It is possible to read off the mean number of quanta $N_{\rm G}$ from the normalisation
condition $\braket{g|g}=1$, which turns out to be
\begin{equation}
N_{\rm G}
=
\bra{g}
\int_0^\infty
\frac{k^2\,\mathrm{d} k}{2\,\pi^2}
\,\hat{a}^\dagger_k\,\hat{a}_k\,
\ket{g}
=
\int_0^\infty
\frac{k^2\,\mathrm{d} k}{2\,\pi^2}
\, g^2_k
\ ,
\end{equation}
and $N_{\rm G}$ is shown to precisely equal the total occupation number of modes in the state $\ket{g}$.
For a uniform source of finite size $R$ and ADM mass $M$, this amount can be estimated
as
\begin{equation}
N_{\rm G}
\sim
\frac{M^2}{m_{\rm p}^2}\,\ln\!\left(\frac{R_\infty}{R}\right)
\ ,
\label{NgR}
\end{equation}
where $R_\infty=k_0^{-1} \gg R$ is a IR cut-off, which cures the divergence coming from
embedding a static (and eternal) field in an infinite space-time.
In reference to that, it is interesting to point out that the dependence of $N_{\rm G}$ 
is weaker on $R$ than on the mass $M$, since
\begin{equation}
\frac{\mathrm{d} N_{\rm G}}{N_{\rm G}}
\sim
2\,\frac{\mathrm{d} M}{M}
-\frac{1}{\ln(R_\infty/R)}\,\frac{\mathrm{d} R}{R}
\ ,
\label{dndelta}
\end{equation}
and allows to recover the opening result $N_{\rm G}\sim M^2/m_{\rm p}^2$ 
when $R_\infty$ is arbitrarily large.
We can now add the non-linearity described by Eq.~\eqref{EOMVn1}, rewritten as
\begin{equation}
\triangle V_{(1)}
=
2\,\frac{l_{\rm p}}{m_{\rm p}}\,
\bra{g}\left(\hat\Phi'\right)^2\ket{g}
\ .
\label{EOMVn1q}
\end{equation}
so that
\begin{equation}
2\,\frac{l_{\rm p}}{m_{\rm p}}
\bra{g}
\left(\hat \Phi'\right)^2
\ket{g}
=
J_g+J_0
\ .
\label{expecPhi2}
\end{equation}
$J_0$ is a diverging contribution coming from the vacuum, which can be removed through
normal ordering in the expectation value.
By means of Eq.~\eqref{gk}, one can immediately see that $J_g$ equals the classical 
expression, that is
\begin{equation}
J_{g}
=
2\,\frac{l_{\rm p}}{m_{\rm p}}
\bra{g}
\left(\hat \Phi'\right)^2
\ket{g}
=
2\,\left(V_{(0)}'\right)^2
\ ,
\label{JgV0}
\end{equation}
for any current sourcing the scalar field. The coherent state $\ket{g}$ is therefore an
appropriate basis for a perturbative analysis in Quantum Field Theory.
In order to refine the result, one shall find a modified coherent state $\ket{g'}$, such that 
\begin{equation}
\sqrt{\frac{l_{\rm p}}{m_{\rm p}}}\,
\bra{g'}\hat \Phi\ket{g'}
\simeq
V_{(0)}
+
q_\Phi\,V_{(1)}
\ ,
\label{expP}
\end{equation}
to first order in $q_\Phi$.
As for the classical potential in Eq.~\eqref{Vexp}, we can expand the quantum state $\ket{g'}$ as
$
\ket{g'}
\simeq
\mathcal{N}
\left(
\ket{g}
+
q_\Phi
\ket{\delta g}
\right)
$,
where $\mathcal{N}$ is a normalisation constant.
After manipulating Eq.~\eqref{expP} with some cumbersome algebra~\cite{Casadio:2017cdv},
we succeed in relating the correction to any eigenvalue, $\delta g_k$, 
to the entire set of $g_p$'s, \emph{i.e.}
\begin{equation}
\frac{\triangle \left({\rm Re}\bra{\delta g} \hat \Phi \ket{g} \right)}
{{\rm Re}\bra{\delta g}\ket{g}}
=
\triangle \bra{g}\hat \Phi\ket{g}
+
\sqrt{\frac{l_{\rm p}}{m_{\rm p}}} \,
\frac{\bra{g}\left(\hat \Phi'\right)^2\ket{g}}
{{\rm Re}\bra{\delta g}\ket{g}}
\ .
\label{eq-schifo}
\end{equation}
Of course, this Equation is very complicated and any attempt to deal with it without some sort 
of approximation is destined to fail. Therefore, we follow the argument outlined in
Refs.~\cite{Casadio:2016zpl,becBH} and pretend that almost the whole set of toy gravitons
belongs to one mode of wavelength $\lambda_{\rm G}\simeq R$~\cite{DvaliGomez}, so that
one can finally identify~\cite{Casadio:2017cdv}
\begin{equation}
\delta g _{\bar k}
\simeq
-l_{\rm p}\,\bar k^{3/2}\,\delta\bar k\,g_{\bar k}^2
\sim
-l_{\rm p}\,\bar k^{5/2}\,g_{\bar k}^2
\ ,
\end{equation}
where $\delta \bar k \sim R^{-1}$.
Considering \emph{e.g.} the point-like source of Ref.~\cite{Casadio:2017cdv}, one obtains
\begin{equation}
\delta g _{\bar k}
\sim
-\frac{l_{\rm p}\,M_0^2}{m_{\rm p}^2\,\bar k^{1/2}}
\sim
\frac{l_{\rm p}\,M_0}{m_{\rm p}\,r_0}\,g_{\bar k}
\sim
\frac{R_{\rm H}}{r_0}\,g_{\bar k}
\ ,
\end{equation}
with the help of a UV cut-off $r_0$, that removes the infinities coming from the vanishing spatial
extension of the source. When the result falls within the range of validity of our approximation,
that is $r_0\ll R_{\rm H}$, we see that the perturbation correctly enjoys the relation 
$\delta g _{\bar k}\ll g _{\bar k}$, which is compatible with the classical result that we want
to extend to the domain of quantum physics.
\section{Conclusions and outlook}
We have constructed an effective quantum theory for the gravitational potential
sourced by a static matter distribution and up to first post-Newtonian order,
by approximating the Einstein-Hilbert action in the weak field and non-relativistic regimes.
The result~\cite{Casadio:2017cdv} implies the maximal packing condition~\eqref{maxpack},
which is a signature feature of the corpuscular BH model of 
Dvali and Gomez~\cite{DvaliGomez4}.
Furthermore, we refined the expression~\eqref{NgR} for the total number of soft quanta,
$N_{\rm G}$, forming the self-sustained BEC and we showed that it is not only influenced 
by the total ADM mass of the black hole as in standard literature~\cite{DvaliSlava},
but also encodes more information, albeit to a much weaker extent. 
In fact, the logarithmic dependence on the ratio $R/R_\infty$ becomes more and more 
negligible as the system approaches an ideal configuration, for which $R_\infty\to\infty$,
while it is expected to give a non-trivial contribution in a dynamical situation.
Moreover, we stated several times that our analysis looses accuracy for a source of size 
$R\lesssim R_{\rm H}$, as we have shown that the consistency relation $\delta g_k \ll g_k$
gets spoiled as one moves further away from it.
Still, it would be interesting to investigate the case where the inequality
is saturated, that is $R=R_{\rm H}$, since Eq.~\eqref{maxpack} arises precisely in this 
regime and allows the black hole to be self-sustained.
The possible outcome may be able to quantify the departure of the present analysis
from the standard post-Newtonian approximation of GR, possibly providing some glimpses
of a full theory of Quantum Gravity.
In addition to that, it is necessary to understand the role of matter pressure, which
has been completely neglected so far, but may have important cosmological
implications~\cite{Casadio:2017twg,Cadoni:2017evg}.
\ack
This proceeding is based on a series of papers in collaboration with R. Casadio, A. Giusti and
M. Lenzi.
The author is partially supported by the ERC Advanced Grant 339169 "Self-completion".
This work has also been carried out in the framework of activities of the National Group 
of Mathematical Physics (GNFM, INdAM).
%
\section*{References}
\end{document}